\documentclass[apj]{emulateapj}

\usepackage{amssymb, amsmath, apjfonts, natbib, color, graphicx}

\newdimen\digitwidth
\setbox1=\hbox{0}
\digitwidth=\wd1
\catcode`"=\active
\def"{\kern\digitwidth}

\slugcomment{Accepted for Publication on ApJ}

\begin{document}

\title{Dissecting the Phase Space Snail Shell}

\author{Zhao-Yu Li\altaffilmark{1, 2, 3} Juntai Shen\altaffilmark{1, 2, 3, 4}}

\altaffiltext{1}{Department of Astronomy, School of Physics and Astronomy, Shanghai
Jiao Tong University, 800 Dongchuan Road, Shanghai 200240, China; correspondence should be addressed to: lizy.astro@sjtu.edu.cn; jtshen@sjtu.edu.cn}
\altaffiltext{2}{Shanghai Key Laboratory for Particle Physics and Cosmology, 200240, Shanghai, China}
\altaffiltext{3}{Shanghai Astronomical Observatory, Chinese Academy of Sciences, 80 Nandan Road, Shanghai 200030, China}
\altaffiltext{4}{College of Astronomy and Space Sciences, University of Chinese Academy of Sciences, 19A Yuquan Road, Beijing 100049, China}

\begin{abstract}

The on-going vertical phase mixing, manifesting itself as a snail shell in the $Z-V_{Z}$ phase space, has been discovered with the Gaia DR2 data. To better understand the origin and properties of the phase mixing process, we study the vertical phase-mixing signatures in arches (including the classical ``moving  groups'') of the $V_{R}-V_{\phi}$ phase space near the Solar circle. Interestingly, the phase space snail shell exists only in the arches with $|V_{\phi} - V_{\rm LSR}| \lesssim 30$ km/s, i.e., stars on dynamically ``colder'' orbits. The snail shell becomes much weaker and eventually disappears for increasingly larger radial action ($J_{R}$), quantifying the ``hotness'' of orbits. Thus one should pay closer attention to the colder orbits in future phase mixing studies. We also confirm that the Hercules stream has two branches (at fast and slow $V_{\phi}$), which may not be explained by a single mechanism, since only the fast branch shows the prominent snail shell feature. The hotter orbits may have phase-wrapped away already due to the much larger dynamical range in radial variation to facilitate faster phase mixing. To explain the lack of a well-defined snail shell in the hotter orbits, the disk should have been perturbed at least $500$ Myr ago. Our results offer more support to the recent satellite-disk encounter scenario than the internal bar buckling perturbation scenario as the origin of the phase space mixing. Origin of the more prominent snail shell in the $V_{\phi}$ color-coded phase space is also discussed.

\end{abstract}


\section{INTRODUCTION}
\label{sec:intro}

The Milky Way disk is not in a complete dynamical equilibrium. It shows prominent structure in kinematic space that is phase mixing in both horizontal and vertical directions. Since the full phase-space information of individual stars may be obtained, the Milky Way is unique and valuable to reveal the disk dynamical evolution in detail. The picture of the Milky Way evolution is complicated, which is affected by both internal and external perturbations. The resonances of the bar and spiral arms can significantly influence the stellar orbits to cause the radial migration in the disk \citep{friedl_etal_94, sel_bin_02, roskar_etal_08, min_fam_10} and to generate substructures in the velocity phase space for stars in the solar neighborhood \citep{dehnen_00, fux_01, antoja_etal_09, antoja_etal_11, quille_etal_11, hun_bov_18}. Large scale bulk motions observed in the Galactic disk \citep[e.g.,][]{sieber_etal_11, carlin_etal_13, sun_etal_15, tian_etal_17, wang_etal_18a, wang_etal_18b} could also be induced from dynamical processes related to the bar and spiral arms \citep{sieber_etal_12, debatt_14, faure_etal_14, monari_etal_15, monari_etal_16}. Moreover, satellite galaxies or sub-halos interacting with the Milky Way can perturb the disk to generate warps, flares or other vertical motions such as the bending and breathing features in the outer disk \citep{hun_too_69, quinn_etal_93, kazant_etal_08, purcel_etal_11, gomez_etal_13, widrow_etal_14, donghi_etal_16, laport_etal_18a, laport_etal_18b}.

Using the revolutionary Gaia data \citet{antoja_etal_18} discovered clear evidence of vertical phase mixing in the solar neighborhood for the first time. A clear snail shell can be seen in the $Z-V_{Z}$ phase space. As first shown in \citet{laport_etal_19}, the phase space snail shell is also prominent in the number density contrast map. This feature can be understood as the phase mixing process in a vertically perturbed disk, where the vertical oscillation frequency depends on the oscillation amplitude, generating a snail shell structure in $Z-V_{Z}$ space \citep{tremai_99, antoja_etal_18}. \citet{candli_etal_14} investigated the evolution of the phase mixing process for disrupting star clusters that show development and winding of a phase space spiral due to the anharmonic oscillation\footnote{See Fig.~4.27 in \citet{bin_tre_08} for a schematic view of phase mixing process.}. In fact, the concept of phase mixing/wrapping has been implied in several previous works. \citet{minche_etal_09} showed that an unrelaxed disk can produce wave-like features in the velocity phase space that get closer due to the phase wrapping. As shown in \citet{quille_etal_09}, external perturbations could excite phase mixing for stars in the disk to induce streams in the velocity distribution. \citet{delave_etal_15} found that phase wrapping excited by external perturbations might account for the bending and breathing modes in the disk. The North/South asymmetry in the vertical stellar number density profile discovered in \citet{widrow_etal_12} is also a reflection of the snail shell with all the stars projected on the $Z$ axis.

The phase space snail shell shows up more clearly when color-coded with azimuthal velocity ($V_{\phi}$) of stars (Fig.~1c in \citealt{antoja_etal_18}), which was suggested to indicate the tight correlation between the in-plane and vertical motions. The two motions are clearly entangled, but there are still important details to be clarified \citep{bin_sch_18, dar_wid_19}. \citet{laport_etal_19} found snail shell at different stellar age bins, and the shape of the snail changes systematically across the Galactic disk. Similar results are also obtained in other studies \citep{tian_etal_18, wang_etal_19}. One possible culprit of the vertical perturbation is the merging Sagittarius dwarf with the last pericentric passage occurred at $\sim 300-900$ Myr ago \citep{antoja_etal_18, bin_sch_18, blandh_etal_19, laport_etal_19}. The other competing scenario is the spontaneous bending waves as a source of long lived internal vertical perturber \citep{che_wid_17}, also including the perturbation from a bar buckling event \citep{khoper_etal_19}. 

It has been well known that the velocity phase space in the solar neighborhood, e.g., $V_{R} - V_{\phi}$, shows complex kinematic substructures, known as ``moving groups'' \citep{dehnen_98, skulja_etal_99, famaey_etal_05, antoja_etal_08}. Gaia revealed new configurations in the phase space, which are the multiple arches in the $V_{R}-V_{\phi}$ space and diagonal ridges in the $R-V_{\phi}$ space 
\citep{gaia_etal_18b, antoja_etal_18}. The arches appear for the whole range of azimuthal velocity, with all the classical moving groups embedded in the more extended arched substructures \citep{gaia_etal_18b}. \citet{ramos_etal_18} remarked that the more roundish moving groups and the elongated arches are morphologically different entities, with some moving groups belonging to the same arch. To the first order, the velocity phase space could be considered as an assembly of arches. Previous studies argued that internal dynamical effects of structures in the Galactic disk mainly generate kinematic features with radial velocity ($U$) and azimuthal velocity ($V$) with respect to the local circular motion within $\sim 50$ km/s \citep{minche_etal_09}. On the other hand, stars with larger radial and azimuthal velocities resembling arc-like features in the $U-V$ plane may be related to the disk ringing effect caused by external interactions with satellite galaxies \citep{minche_etal_09, gomez_etal_12}.

Considering the different internal or external origins of the arches, the connection between the arches and previous vertical perturbation events is still an open question. For example, \citet{michtc_etal_19} reported that the snail shell is only produced by the classical moving groups, with no evidence of incomplete vertical phase mixing from external perturbations. This apparent inconsistency with other works needs to be better understood by properly dissecting the velocity phase space into distinct arches. Here we utilize the second Gaia data release (DR2) with radial velocity, proper motions and parallax available to investigate the properties of different arches in the $Z-V_{Z}$ phase space. Hopefully, this study will shed light on the origin of the snail shell and the correlation between the in-plane and vertical motions. The sample is described in \S 2. The results are shown and discussed in \S 3, and summarized in \S 4.

\begin{figure}
\center
\includegraphics[scale=1.9, viewport = 15 5 120 130]{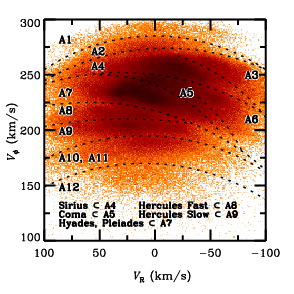}
\caption{The $V_{R}-V_{\phi}$ phase space of the sample around $R = 8.34$ kpc. Arches are separated with the dotted lines. Table~\ref{tab:sample} also lists the classical moving groups and the corresponding arches that they belong to.}
\label{fig:uv}
\end{figure}

\begin{deluxetable*}{llcccc} 
\tabletypesize{\footnotesize} 
\tablecolumns{6} 
\tablewidth{0pt} 
\tablecaption{Properties of the Arches in the $V_{R}-V_{\phi}$ Phase Space} 
\tablehead{
\colhead{Arch ID} & \colhead{Contains} &  \colhead{Number of Stars} & \colhead{Snail Amplitude} & \colhead{Snail Prominence}  & \colhead{Range of $Z$ slit}\\
\colhead{} & \colhead{} & \colhead{} & \colhead{} & \colhead{} & \colhead{(kpc)}\\
\colhead{(1)} & \colhead{(2)} & \colhead{(3)} & \colhead{(4)} & \colhead{(5)} & \colhead{(6)}}
\startdata 

A1 & $-$ & 8,966 & 0.086 & Indistinct &  $[-0.34, -0.20]$ \\
A2 & $-$ & 16,335 & 0.005 & Indistinct & $[-0.34, -0.20]$  \\
A3 & $\gamma$Leo & 36,540 & 0.422 & Moderate &  $[-0.32, -0.24]$ \\
A4 & Sirius & 183,689 & 0.332 & Prominent &  $[-0.34, -0.28]$ \\
A5 & Coma, Dehnen98$-$6 & 106,166 & 0.282 & Prominent &  [-0.30, -0.20] \\
A6 & Dehnen98$-$14 & 61,412 & 0.186 & Moderate &  [-0.26, -0.18] \\
A7 & Hyades, Pleiades & 287,909 & 0.304 & Prominent &  [-0.26, -0.18] \\
A8 & Hercules Fast & 58,537 & 0.377 & Prominent &  [-0.34, -0.26] \\
A9 & Hercules Slow, $\epsilon$Ind & 84,978 & 0.045 & Indistinct &  $[-0.40, -0.28]$ \\
A10, A11 & HR 1614, Bobylev16$-$22 & 49,909 & 0.027 & Indistinct &  $[-0.40, -0.28]$ \\
A12 & Arcturus & 17,383 & 0.059 & Indistinct &  $[-0.40, -0.28]$ \\

 \enddata 
\tablecomments{Col. (1): Arch ID adopted from \citet{ramos_etal_18}. Col. (2): Classical moving groups contained in the arch. Col. (3): Number of stars in this arch. Col. (4): Amplitude/contrast of the snail shell in each arch, which is defined as the difference between the local maximum and minimum values in the corresponding $V_{Z}$ ranges of the $\Delta N$ profile along a narrow $Z$ slit in the phase space. Col. (5): Prominence of the phase space snail shell. Col. (6): The range of $Z$ slit used to extract the $\Delta N$ profile for the snail shell amplitude calculation.}
\label{tab:sample}
\end{deluxetable*}

\begin{figure*}
\epsscale{1.2}
\plotone{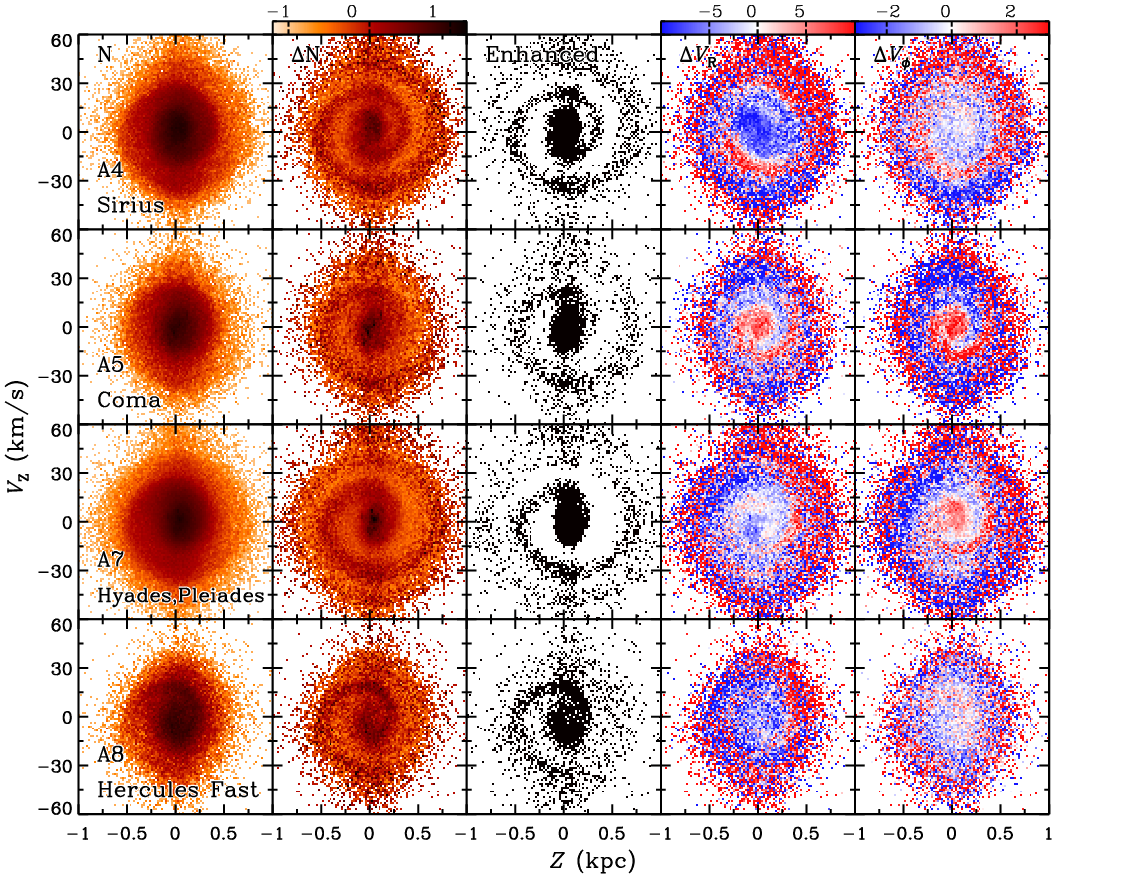}
\caption{The $Z-V_{Z}$ phase space of arches A4, A5, A7 and A8 showing prominent snail shell pattern. From left to right, the columns correspond to the phase space distribution color-coded with number density $N$, number density contrast $\Delta N$, density enhanced map, radial velocity $V_{R}$ and azimuthal velocity $V_{\phi}$, respectively. Each pixel corresponds to $0.02\ {\rm kpc} \times 1.2\ {\rm km/s}$. The Gaussian kernel width used to construct the density contrast map is empirically chosen as 14 pixels to best highlight the snail shell. Note that in the $V_{R}$ and $V_{\phi}$ color-coded phase space, the median value has been subtracted.}
\epsscale{1.0}
\label{fig:snail_cle}
\end{figure*}

\section{SAMPLE}
\label{sec:samp}

\subsection{Sample Selection}

Gaia DR2 opens a new era of precise stellar dynamics, providing astrometric parameters for 1.3 billion sources down to $G \sim 21$ mag, as well as line-of-sight velocities for 7.2 million stars brighter than 12 mag, with the median parallax uncertainty for bright sources ($G < 14$ mag) at 0.03 mas and the proper motion uncertainty at 0.07 mas/yr \citep{gaia_etal_18a}. We adopted the same sample selection as \citet{antoja_etal_18}, i.e., selecting stars with positive parallaxes $\varpi$ with relative uncertainty less than 20\% ($\varpi / \sigma_{\varpi} > 5$). As pointed out by \citet{antoja_etal_18}, the sample selection makes $1/\varpi$ a reasonably good distance estimator \citep{sch_aum_17, luri_etal_18}.

The sample consists of 6.2 million stars, covering the region with $4 < R < 13$ kpc. Following \citet{antoja_etal_18}, we adopt $(X_{\odot},\ Y_{\odot},\ Z_{\odot})$ = ($-$8.34, 0, 0.027) kpc as the Sun position \citep{reid_etal_14}. The local standard of rest (LSR) circular velocity $V_{\rm LSR}$ is set to 240 km/s \citep{reid_etal_14}. Here we adopt the peculiar velocities of the Sun with respect to LSR as $(U_{\odot}, V_{\odot}, W_{\odot})$ = (11.1, 12.24, 7.25) km/s \citep{schonr_12}. Our main results are not affected if we choose other measurements of the solar peculiar motion, e.g., \citet{tian_etal_15} or \citet{huang_etal_15}.  The typical velocity uncertainty is about 1 km/s for the radial, azimuthal, and vertical velocities \citep{gaia_etal_18a, antoja_etal_18}. 

In this study, we select the stars in a narrow annulus in the solar neighborhood ($R = 8.34 \pm 0.1$ kpc), which contains $\sim$ 0.93 million stars. The $V_{R}-V_{\phi}$ phase space distribution of stars in the solar neighborhood is known to show a variety of arches and clumps. The position of the major arches are consistent with recent works \citep{gaia_etal_18b, antoja_etal_18, ramos_etal_18}. We confirm that the stars in this sample show the same snail shells in the $Z-V_{Z}$ phase space when color-coded with the number density, $V_{R}$, and $V_{\phi}$ as in \citet{antoja_etal_18}. 

To evaluate the influence of the parallax bias in the Gaia catalog, we also tested our results with the parallax corrected Gaia sample \citep{schonr_etal_19}. The parallax corrected sample size is reduced to 60\% of that used here. The snail shell patterns in both samples are in excellent agreement; our results and main conclusions are unaffected by the parallax correction, since most of the stars are still very close to the solar neighborhood.

\subsection{Identification of Arches}

We further dissect the sample in the $V_{R}-V_{\phi}$ phase space into different arches containing the classical moving groups, e.g., Sirius, Hyades, Coma Berenices (hereafter Coma for brevity), Pleiades, and Hercules. By applying the Stationary Wavelet Transform \citep{sta_mur_02} on the $V_{R}-V_{\phi}$ distribution of Gaia DR2 data in the solar neighborhood, \citet{ramos_etal_18} identified 12 arches (A1 to A12). Adopting the positions and extensions of the arches in \citet{ramos_etal_18}, we dissect the velocity phase space into different regions corresponding to these 12 arches. 

We identify the gaps between these arches by adjusting the contrast level of the number density map (see Appendix for more information). Fig.~\ref{fig:uv} shows the arches separated by the dotted lines in the velocity phase space. Clearly, the arches are well separated.\footnote{Arches A10 and A11 are grouped together due to the ambient boundary between them.} The classical moving groups, such as the Sirius, Coma and Hyades-Pleiades, are embedded in arches A4, A5, and A7, respectively. Note that arches A8 and A9 together form the Hercules stream \citep{gaia_etal_18b, ramos_etal_18}. Considering their azimuthal velocity difference, we refer to A8 as the Hercules Fast branch with median $V_{\phi} \sim 209$ km/s, and A9 as the Hercules Slow branch with median $V_{\phi} \sim 202$ km/s. Properties of the arches are listed in Table~\ref{tab:sample}. $V_{Z}$ is not considered when classifying arches.
 
To highlight the snail shell in the number density map of the $Z-V_{Z}$ phase space, we adopt the method in \citet{laport_etal_19} to derive the number density contrast $\Delta N$,
\begin{equation}
\Delta N = \frac{N}{\widetilde{N}} - 1,
\end{equation}
where $\widetilde{N}$ is the Gaussian kernel convolved number density distribution.

To further increase the visibility of the snail shell, we generate the enhanced map based on the $\Delta N$ map. Since the inter-shell region in the $\Delta N$ map has values typically less than $-0.03$, we can significantly enhance the snail shell pattern by showing only those regions with $\Delta N \geq -0.03$.

\begin{figure*}
\epsscale{1.2}
\plotone{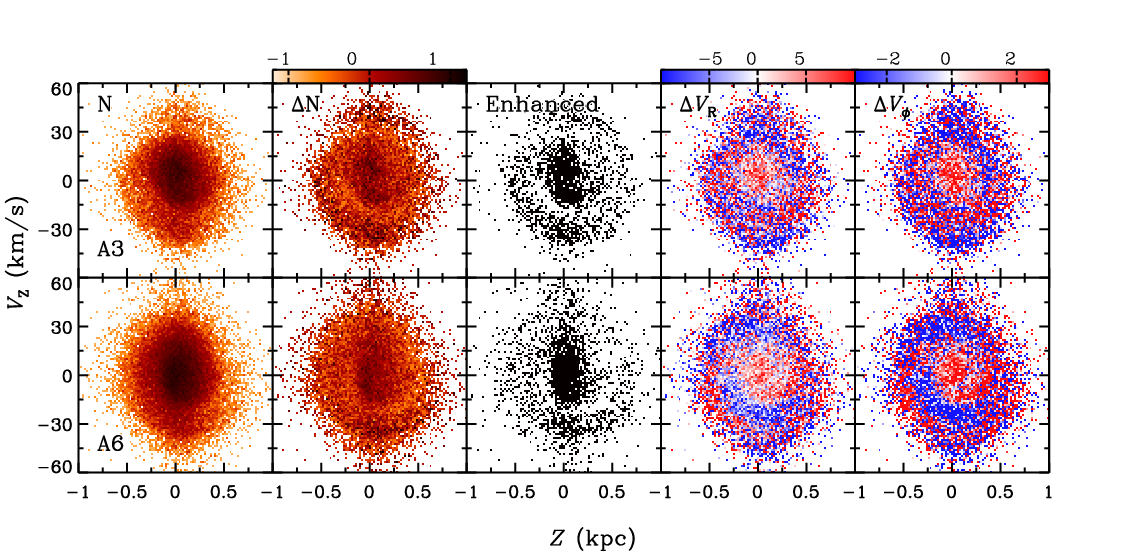}
\caption{The $Z-V_{Z}$ phase space of arches A3 and A6 showing moderate snail shell feature. The layout is the same with  Fig.~\ref{fig:snail_cle}. The prominence of the snail shells are moderate here.}
\epsscale{1.0}
\label{fig:snail_mod}
\end{figure*}

\begin{figure*}
\epsscale{1.2}
\plotone{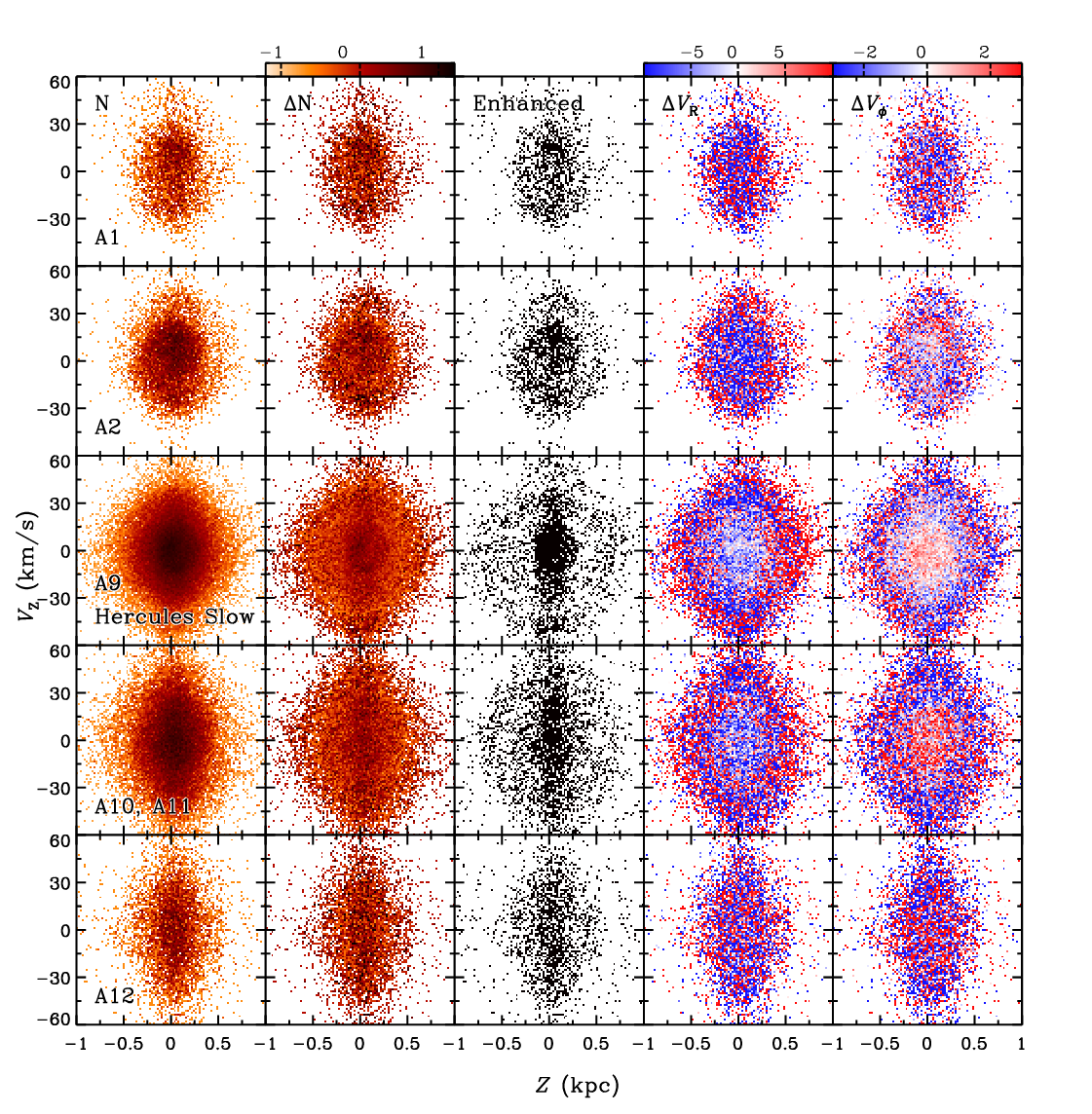}
\caption{The $Z-V_{Z}$ phase space of arches A1, A2, A9, A10, A11 and A12 showing blurred or absent snail shell feature. The layout is the same with Fig.~\ref{fig:snail_cle}. The snail shells here are indistinct compared to Figs.~\ref{fig:snail_cle} and ~\ref{fig:snail_mod}.}
\epsscale{1.0}
\label{fig:snail_no}
\end{figure*}

\begin{figure}
\epsscale{1.2}
\plotone{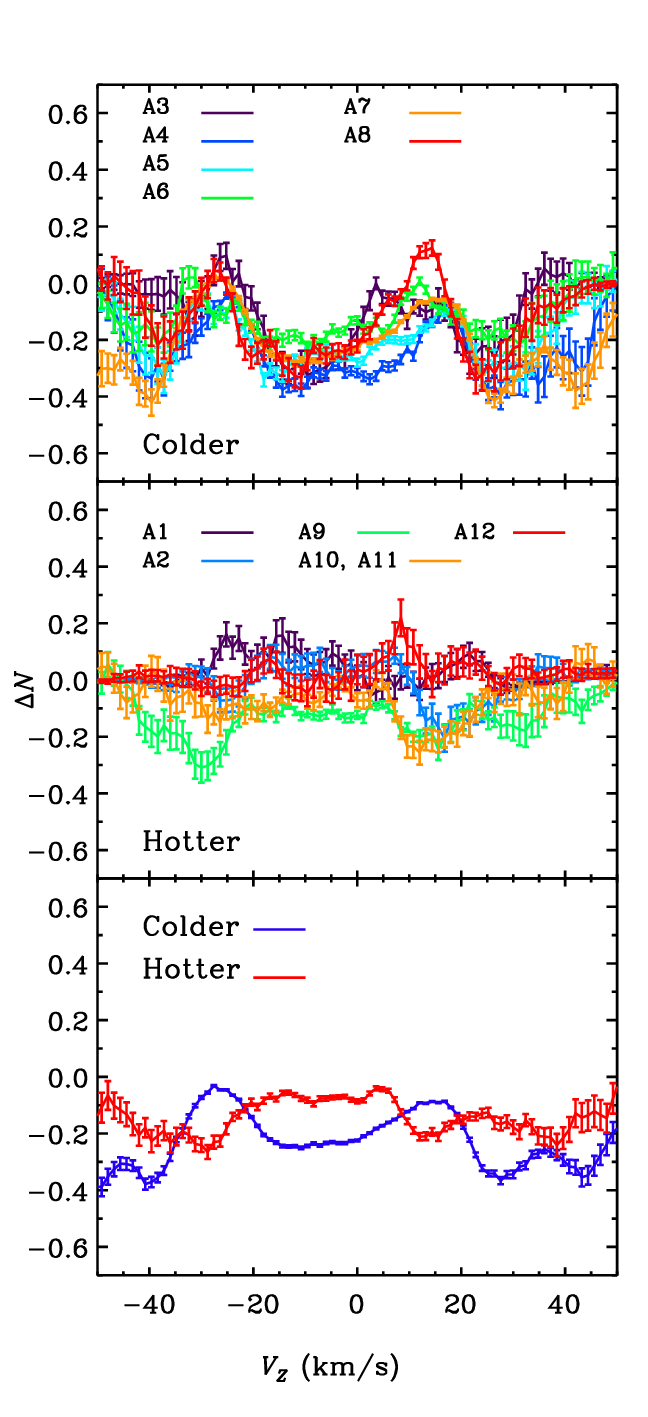}
\caption{Snail shells amplitude quantification. Number density contrast profiles along a constant $Z$ slit (see Table~1) for arches in the colder and hotter orbits are shown in the top and middle panels, respectively. The bottom panel compares the density profiles of all the stars on colder or hotter orbits; from the profiles, we use the difference between the local maximum within $-30 < V_{Z} < -20$ and the local minimum within $-20 < V_{Z} < 10$ to quantify the amplitude of the snail shell.}
\epsscale{1.0}
\label{fig:prof}
\end{figure}

\begin{figure}
\epsscale{1.2}
\plotone{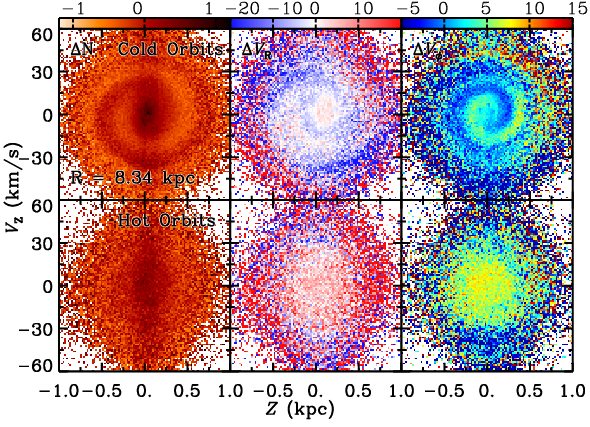}
\caption{The $Z-V_{Z}$ phase space of all the stars on the colder orbits (top row) and hotter  orbits (bottom row). The colder orbits include arches A3, A4, A5, A6, A7, and A8, while the hotter orbits contain A1, A2, A9, A10, A11 and A12. The left, middle, and right columns show the phase space color-coded with $\Delta N$, $\Delta V_{R}$, and $\Delta V_{\phi}$, respectively.}
\epsscale{1.0}
\label{fig:cold_hot}
\end{figure}

\begin{figure}
\epsscale{1.2}
\plotone{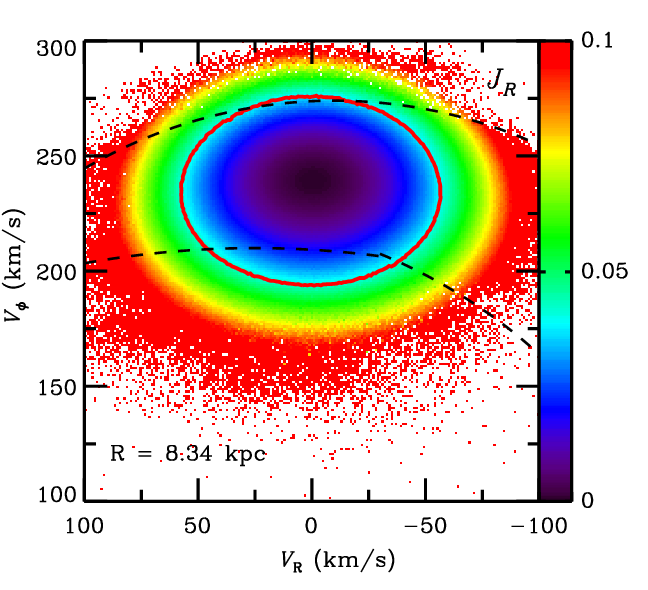}
\caption{The $V_{R}-V_{\phi}$ phase space distribution of our sample color-coded with $J_{R}$, which quantifies the extent of the radial oscillation. $J_{R}$ has the dimension of the angular momentum in the unit of ${\rm kpc^{2}Myr^{-1}}$. The red ellipse represents the contour at $J_{R} = 0.04$ to separate the colder and hotter orbits. The black dashed lines mark the boundaries in Fig.~\ref{fig:uv} for the colder and hotter orbits. They are in agreement in terms of the range of $V_{\phi}$.}
\epsscale{1.0}
\label{fig:vrvp_jr}
\end{figure}

\section{Results and Discussion}
\label{sec:results}

\subsection{Dissecting the Phase Space Snail Shell}

The $Z-V_{Z}$ phase space distributions of the arches are shown in Figs.~\ref{fig:snail_cle},~\ref{fig:snail_mod}, and~\ref{fig:snail_no}, which are color-coded with number density $N$, number density contrast $\Delta N$, enhanced map, radial velocity $V_{R}$ and azimuthal velocity $V_{\phi}$. The arches are grouped according to the prominence of the phase space snail shell. The arches with prominent snail shells are shown in Fig.~\ref{fig:snail_cle}, arches with moderate snail shells in Fig.~\ref{fig:snail_mod}, and arches with indistinct snail shells in Fig.~\ref{fig:snail_no}.

In Fig.~\ref{fig:snail_cle}, for each arch, the snail shells revealed in $N$, $\Delta N$ and enhanced maps are almost identical. The snail shell pattern is slightly different in the $V_{R}$ and $V_{\phi}$ color-coded phase space compared to the number density map, but the amplitude contrast of the snail shell is only $2-4$ km/s. This may be due to the variation of the snail shell shape with the radial or azimuthal velocities for each arch, which will be discussed in greater detail in Section 3.5. For arch A8, there seems to be no visible snail shell in the $V_{R}$ or $V_{\phi}$ color-coded phase space.

From the $V_{R}-V_{\phi}$ phase space in Fig.~\ref{fig:uv}, it seems that the snail shell only exists for arches with $210 \lesssim V_{\phi} \lesssim 270$ km/s (i.e., $|V_{\phi} - V_{\rm LSR}| \lesssim 30$ km/s). This velocity range corresponds to the dynamically colder orbits, which are closer to circular orbits. Stars with velocities outside that region are denoted as hotter orbits. 

We also quantify the amplitude/contrast of the snail shell. For the $Z-V_{Z}$ phase space distribution of each arch, along a constant $Z$ slit, $\Delta N$ profile as a function of $V_{Z}$ is extracted. The $Z$ slit range listed in Table~\ref{tab:sample} is purposely chosen to cut through the shell (local maximum) and inter-shell regions (local minimum). Profiles of the number density contrast are shown in Fig.~\ref{fig:prof}. The colder orbits (top panel) show large fluctuations with a similar zigzag pattern due to the multiple intersections of the snail shell with the slit, while the hotter orbits (middle panel) show roughly flat profiles with small fluctuations, indicating the weak or indistinct snail shell in the hotter orbits. To quantify the amplitude of the snail shell, in each profile, we measure the difference between the maximum value within $-30 < V_{Z} < -20$ km/s and the minimum value within $-20 < V_{Z} < 10$ km/s. The results are also listed in Table~\ref{tab:sample}. Clearly, the amplitude/contrast of the snail shell in colder orbits ($\sim 0.3$) is significantly higher than that of the hotter orbits ($\sim 0.03$).

Fig.~\ref{fig:cold_hot} shows the phase space distributions for all the stars on colder orbits (top row; including A3, A4, A5, A6, A7, and A8) and hotter orbits (bottom row; including A1, A2, A9, A10, A11, and A12). Stars on the colder orbits clearly show a very prominent snail shell in the number density (left) and $V_{\phi}$ color-coded phase space (right), and a slightly weaker snail shell in the $V_{R}$ color-coded phase space (middle). On the other hand, the hotter orbits (the bottom row) do not exhibit the clear snail shells in the number density, $V_{R}$ or $V_{\phi}$ color-coded phase spaces. If the phase space snail shell really exists in the hotter orbits, it is significantly blurred or incoherent. The bottom panel in Fig.~\ref{fig:prof} compares the density profiles between the combined colder and hotter orbits, confirming our argument for the prominent snail shell in the colder orbits only. Notice that in Fig.~\ref{fig:cold_hot}, for the colder orbits, the snail shell patterns between the $\Delta N$ and $V_{\phi}$ color-coded phase spaces are different. This probably indicates that the snail shell shape varies across the arches in the colder orbits. This will be discussed in greater detail in Section 3.5.

As shown in Fig.~\ref{fig:snail_no}, the phase space distributions of arches A10, A11 and A12 are more elongated along the $V_{Z}$ axis than arches A1 and A2. This is consistent with simple theoretical expectation of approximately harmonic oscillators. The difference between the median azimuthal velocities of the arches at high and low $V_{\phi}$ is $\sim$ 100 km/s, corresponding to $\sim$ 3.5 kpc in terms of the difference in the guiding radius ($R_{g}$)\footnote{$R_{g} \approx \frac{V_{\phi} \times R_{0}}{V_{c}}$, assuming a flat rotation curve.}. Therefore, compared to the arch at higher $V_{\phi}$, stars with lower $V_{\phi}$ have much smaller $R_{g}$, where both the vertical oscillation frequency and vertical velocity amplitude are larger ($V_{\rm Z, max} \sim \Omega_{\rm z} Z_{\rm max}$), resulting in a more elongated distribution along the vertical velocity axis in the $Z-V_{Z}$ phase space. This simple argument is also consistent with the snail shape variation found by \citet{laport_etal_19} in their Fig. 15 for a region centered on $R = 6$ kpc, which is more elongated along the $V_{Z}$ axis compared to the snail shell at $R = 10$ kpc. Similar results are also seen in \citet{wang_etal_19}.

\begin{figure*}
\epsscale{1.}
\plotone{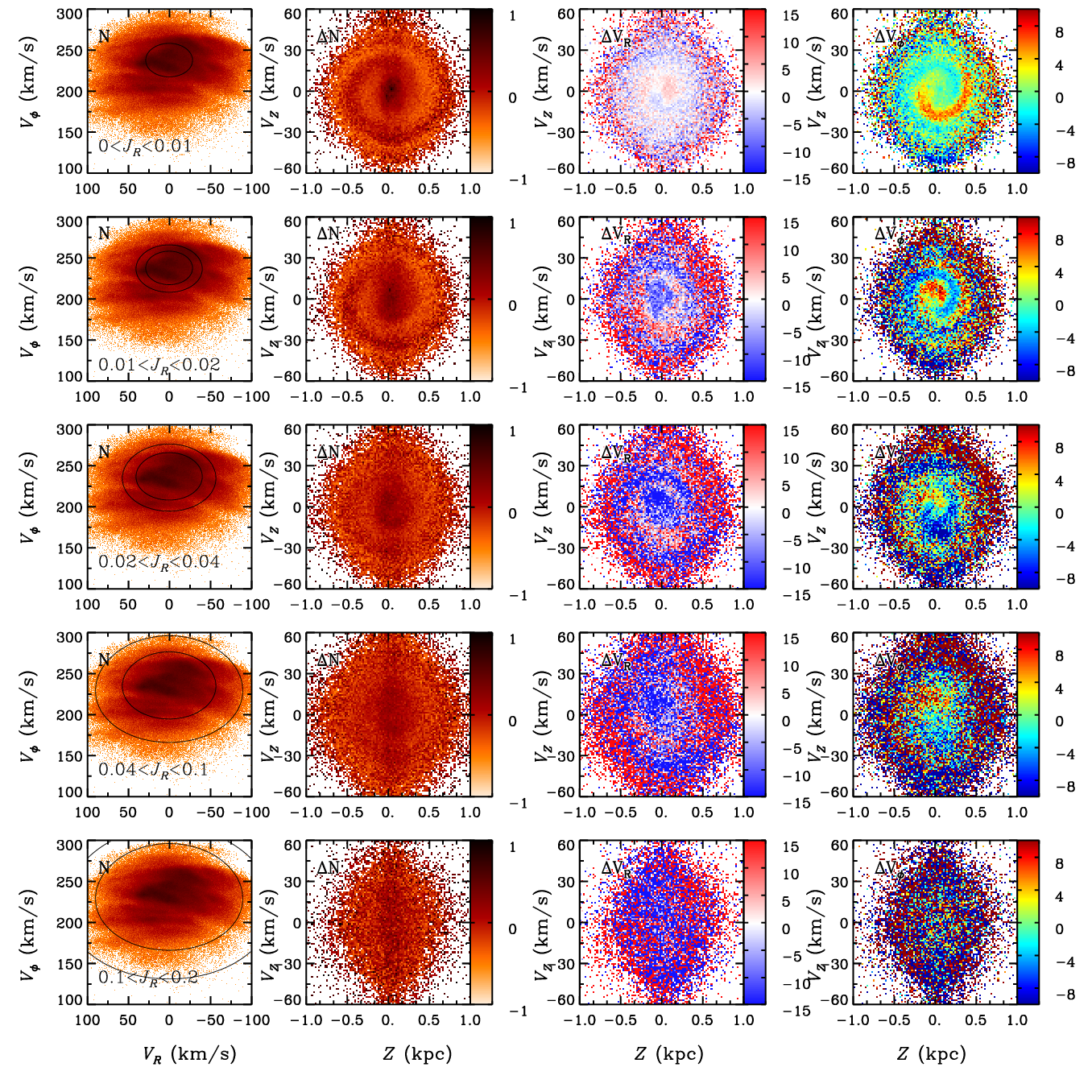}
\caption{$Z-V_{Z}$ phase space distributions for stars in different $J_{R}$ ranges. The left column shows  $V_{R} - V_{\phi}$ phase space of the whole sample overlaid with ellipses contours of the lower and upper boundaries of the $J_{R}$ ranges. For each $J_{R}$ range, the second, third and fourth columns represent the $Z-V_{Z}$ phase space distributions color-coded with $\Delta N$, $V_{R}$ and $V_{\phi}$, respectively. From top to bottom rows, the gradual disappearance of a snail shell in increasingly ``hotter'' (higher $J_{R}$) orbits is clear.}
\epsscale{1.0}
\label{fig:zvz_jr}
\end{figure*}

\begin{figure}
\epsscale{1.2}
\plotone{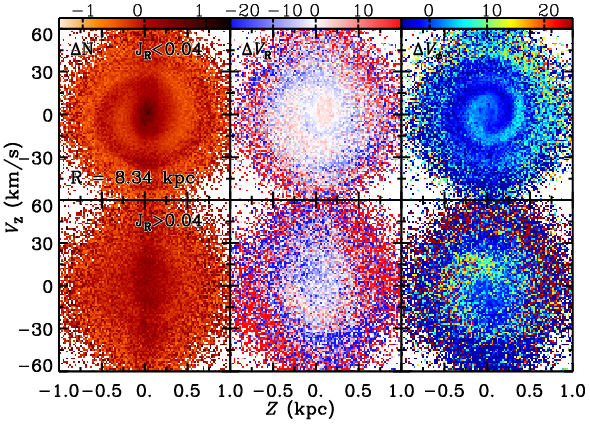}
\caption{The $Z-V_{Z}$ phase space distributions for the stars with $J_{R} < 0.04$ (colder orbits; top row) and $J_{R} > 0.04$ (hotter orbits; bottom row). The left, middle, and right columns show the phase space color-coded with $\Delta N$, $V_{R}$, and $V_{\phi}$, respectively.}
\epsscale{1.0}
\label{fig:cold_hot_jr}
\end{figure}

One may use the radial action $J_{R}$ to quantify the extent of the radial oscillation (or ``hotness'') of orbits. We use the action-based galaxy modeling package AGAMA \citep{vasili_19} to directly compute $J_{R}$ of each star in our sample, with the best-fit potential from \citet{mcmill_11} adopted. $J_{R}$ has the dimension of angular momentum. Its unit in our paper is ${\rm kpc^{2}Myr^{-1}}$. Fig.~\ref{fig:vrvp_jr} shows the $J_{R}$ color-coded $V_{R} - V_{\phi}$ phase space distribution of our sample. As expected, $J_{R}$ is smaller for nearly circular orbits with velocity closer to the circular velocity of the LSR. Fig.~\ref{fig:zvz_jr} shows the $Z-V_{Z}$ phase space distributions for stars in different $J_{R}$ ranges. The snail shell becomes much weaker for increasingly larger $J_{R}$ values; the snail shell is indistinct with $J_{R} > 0.04$. This phenomenon is expected since stars with large $J_{R}$ occupy a wide range of guiding radius distribution, resulting in a significantly blurred  snail shell. Moreover, hotter stars respond poorly to dynamical perturbations.

This could also explain why the snail shells of arches A3 and A6 shown in Fig.~\ref{fig:snail_mod} are not as prominent as those in Fig.~\ref{fig:snail_cle} (for A4, A5, A7, and A8). Stars in A3 and A6 have large radial action ($J_{R} > 0.04$) to blur the snail shell feature.

In Fig.~\ref{fig:vrvp_jr}, the red ellipse represents the contour of $J_{R} = 0.04$, which agrees well with the azimuthal velocity ranges of colder orbits marked with the two black dashed lines. The $Z-V_{Z}$ phase space distributions of stars with $J_{R} < 0.04$ and $J_{R} > 0.04$ are shown in Fig.~\ref{fig:cold_hot_jr}. Clear snail shell can only be seen for the subsample with $J_{R} < 0.04$. The lower panels with $J_{R} > 0.04$ show a significantly blurred or incoherent snail shell feature. This agrees with Fig.~\ref{fig:cold_hot}, confirming that hotter orbits (with larger radial oscillation) do not show prominent snail shell in the phase space. Arches A3 and A6 are partially included in the sample with $J_{R} > 0.04$ to result in a blurred snail shell feature in the lower right panel of Fig.~\ref{fig:cold_hot_jr} for the $V_{\phi}$ color-coded phase space.

The median $J_{R}$ of our sample is 0.018, which is very similar to the result in \citet{blandh_etal_19} within a few percent ($\sim 0.0195$ after being converted to our unit). The larger $J_{R}$ subsample (above the median $J_{R}$) of \citet{blandh_etal_19} shows a weak phase space snail shell. Apparently, their high $J_{R}$ subsample contains a significant fraction of colder orbits defined by us, thus giving rise to the weak snail shell.

\subsection{Two Branches of the Hercules Stream}

The Hercules stream, as a prominent low azimuthal velocity structure in the velocity phase space, has been extensively studied in the literature. Stellar spectra of individual stars in the stream suggest multiple stellar populations with different ages and metallicities, arguing against the stream as a disrupted stellar cluster or the debris of a satellite galaxy \citep{famaey_etal_05, bensby_etal_07}. Previous theoretical works have suggested that the Hercules stream is due to the dynamical effects of the bar and/or spiral arms \citep{dehnen_00, antoja_etal_14, perezv_etal_17, hun_bov_18}. 

Gaia DR2 showed that the Hercules stream is composed of two branches at different azimuthal velocities \citep{gaia_etal_18b, ramos_etal_18}. For the two branches of the Hercules stream, it is surprising to see that only the fast branch (arch A8; bottom row in Fig.~\ref{fig:snail_cle}) shows the prominent snail shell but not for the slow branch (arch A9; middle row in Fig.~\ref{fig:snail_no}). This difference should be physical and not due to small number statistics, since the slow branch (A9) even has $\sim 40$\% more stars than the fast branch (A8) as listed in Table~\ref{tab:sample}. Therefore, the Hercules stream may not be a homogeneous kinematic structure. If the two branches can be explained by a single mechanism, then it is difficult to understand the difference in the vertical phase space distributions, considering the small azimuthal velocity difference ($\sim 7$ km/s) between the two branches.

\begin{figure*}
\plotone{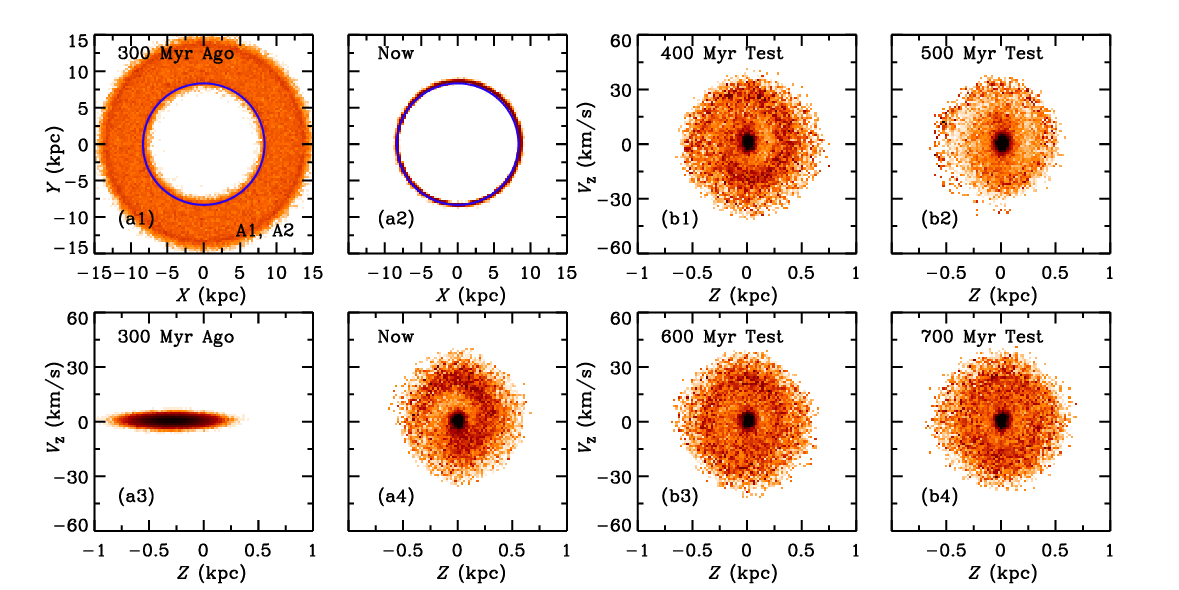}
\caption{Test particle simulation of the first approach showing the phase mixing of stars on hotter orbits in arches A1 and A2. Panels (a1) $-$ (a4) show the simulation results with the vertical perturbation imposed 300 Myr ago. Panels (a1) and (a2) show the face-on distribution of the test particles 300 Myr ago and now, with the blue circle representing the solar radius ($R = 8.34$ kpc). Panels (a3) and (a4) are the corresponding number density contrast map of the $Z-V_{Z}$ phase space. Other test results with the vertical perturbation imposed on the test particle simulations at 400, 500, 600, and 700 Myr ago are shown in panels (b1) - (b4).}
\label{fig:test1}
\end{figure*}

\subsection{Phase Mixing in Colder and Hotter Orbits}

The snail shell in the $Z-V_{Z}$ phase space reflects the vertical phase mixing with anharmonic oscillation \citep{antoja_etal_18}. Stars with larger vertical action $J_{Z}$ have smaller vertical oscillation frequency $\Omega_{Z}$ \citep{bin_sch_18}. \citet{bin_sch_18} also showed that $\Omega_{Z}$ depends sensitively on $V_{\phi}$, with smaller $\Omega_{Z}$ at larger $V_{\phi}$, which forms a narrow and sequential distribution in $\Omega_{Z} - \sqrt{J_{Z}}$ plane for each $V_{\phi}$; the snail shell shape may change slightly at different $V_{\phi}$. Our result shows clear phase space snail shells for stars in individual arch on the colder orbits. The snail shell shapes of these arches are also slightly different from each other as shown in Fig.~\ref{fig:snail_cle}. This is roughly consistent with \citet{bin_sch_18}; each arch usually has a small $V_{\phi}$ range, which corresponds nicely to a narrow strip in the $\Omega_{Z} - \sqrt{J_{Z}}$ plane to induce a clear phase space snail shell.

\citet{bin_sch_18} suggested that the unclear snail shell in the phase space number density distribution observed in \citet{antoja_etal_18} is due to the fact that stars lie in a ``broad swath'' in the $\Omega_{Z} - \sqrt{J_{Z}}$ plane. As shown in the top left panel of Fig.~\ref{fig:cold_hot}, all the stars on colder orbits combined together show prominent snail shell in the phase space number density distribution. These results may indicate that the distribution of stars on colder orbits in the $\Omega_{Z} - \sqrt{J_{Z}}$ plane is more well-defined and narrower than ``a broad swath''.

Both the analytical estimation and simple test particle simulations using {\it galpy\footnote{http://github.com/jobovy/galpy}} \citep{bovy_15} with the MWPotential2014 show that stars on hotter orbits typically have much larger dynamical range in the disk with the radial oscillation amplitudes $(R_{\rm max} - R_{\rm min}) / 2$ $\sim 2$ kpc than stars on colder orbits ($\lesssim 0.7$ kpc). It is therefore natural that hotter orbits will scramble the signal, as hotter stars arrive at this very local sample from a range of guiding radii. As shown in \citet{blandh_etal_19}, given the same amplitude of the vertical perturbation, stars on circular orbits at larger radius will form a more loosely winding phase space snail shell at lower vertical oscillation frequency than the stars on circular orbits at smaller radius. The snail shell at larger radius becomes more elongated in the $Z$ direction, but reduced in the $V_{Z}$ direction ($\Omega_{z} \sim V_{\rm Z,max}/Z_{\rm max}$). Due to the large radial range of the hotter orbits, the elongation of the $Z-V_{Z}$ phase space ellipse changes during its radial oscillation, leading to a blurred distribution and faster phase mixing. 

Stars on the hotter orbits have probably phase-wrapped away already to show significantly blurred and indistinct snail shell, while the stars on the colder orbits are still in the process of vertical phase mixing. To demonstrate this argument we perform a simple simulation with 100,000 test particles in a realistic Milky Way potential \citep{irrgan_etal_13, antoja_etal_18} to track the evolution of the hotter and colder orbits. Note that our test particle simulation is designed to mimic an external vertical perturbation. We have verified that it can well reproduce the simulation results in \citet{antoja_etal_18}, e.g., their Fig. 3a and Extended Figs. 3a and 3b. 

To consider the fact that the perturbation happened in the past, and we observe the stars in the solar neighborhood at present, we retro-tracked the orbits to make sure their present location is indeed near the solar radius. Initially all test particles are distributed around R = 8.34 kpc with median $V_{\phi} = 280$ km/s ($\sigma_{\phi} = 15$ km/s) and median $V_{R} = 0$ km/s ($\sigma_{R} = 20$ km/s), representing arches A1 and A2 at higher $V_{\phi}$. Then the particle orbits are integrated backwards without vertical perturbation ($V_{Z} = 0, Z = 0$). The distributions of the test particles in $X-Y$ plane at 300 Myr ago is shown in panel (a1) of Fig.~\ref{fig:test1}. Then we impose the vertical perturbation on all the particles in two approaches. 

The first approach is similar to \citet{antoja_etal_18}, where the test particles are displaced vertically (median $Z = -0.3$ kpc and $V_{Z} = 0$ km/s with dispersions of 0.2 kpc and 2 km/s). For the 300 Myr case, after the perturbation, the $Z-V_{Z}$ phase space distribution is shown in panel (a3) of Fig.~\ref{fig:test1}. After 300 Myr evolution, almost all the test particles now arrive in the solar radius as shown in panel (a2) in the $X-Y$ plane, with a snail shell feature in the number density contrast map ($\Delta N$) of the $Z-V_{Z}$ phase space distribution in panel (a4). In addition, we test to impose the same vertical perturbation on test particles at 400, 500, 600, and 700 Myr ago. The number density contrast map ($\Delta N$) of the final $Z-V_{Z}$ phase space distributions of the four tests are shown in Panels (b1)$-$(b4) in Fig.~\ref{fig:test1}.

\begin{figure*}
\epsscale{1.}
\plotone{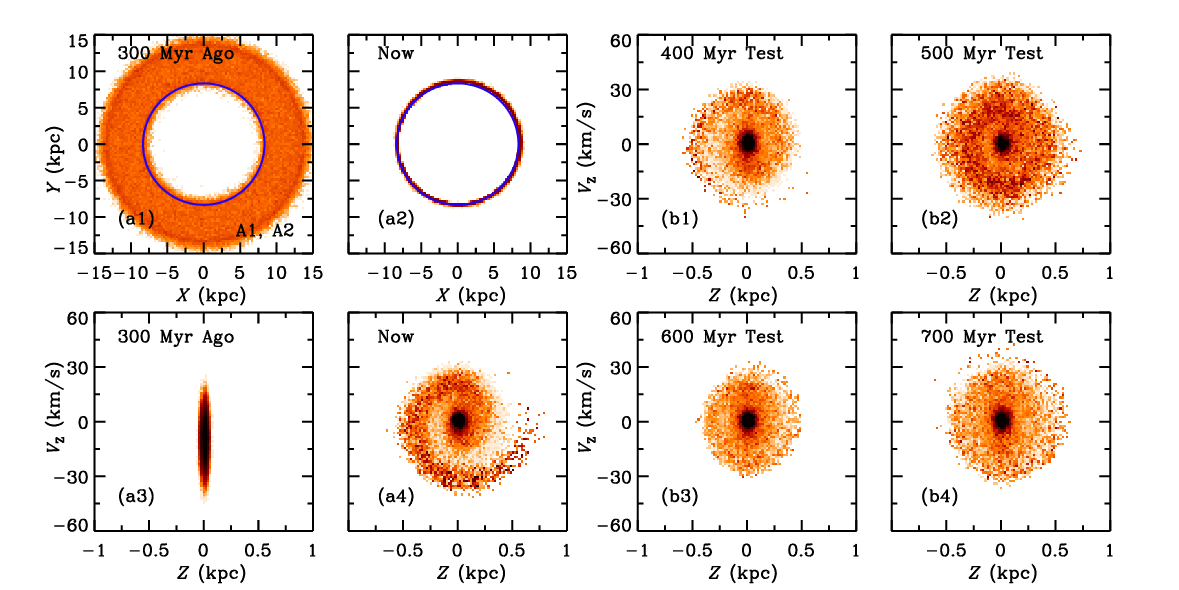}
\caption{Test particle simulation of the second approach showing the phase mixing of stars on hotter orbits in arches A1 and A2. The layout here is the same with Fig.~\ref{fig:test1}.}
\epsscale{1.0}
\label{fig:test2}
\end{figure*}

In the second approach, under the impulsive approximation of the external perturbation \citep{bin_sch_18}, a vertical velocity kick is imposed to all the test particles with the vertical positions barely changed (median $Z = 0$ and $V_{Z} = -10$ km/s with dispersion of 0.02 kpc and 10 km/s). Panels (a1) $-$ (a4) and (b1) $-$ (b4) in Fig.~\ref{fig:test2} show the test particle simulation results. Combining Figs.~\ref{fig:test1} and~\ref{fig:test2}, it seems that the snail shell becomes blurred or indistinct if the perturbation was imposed at least $\sim$ 500 Myr ago. 

\begin{figure*}
\epsscale{1.}
\plotone{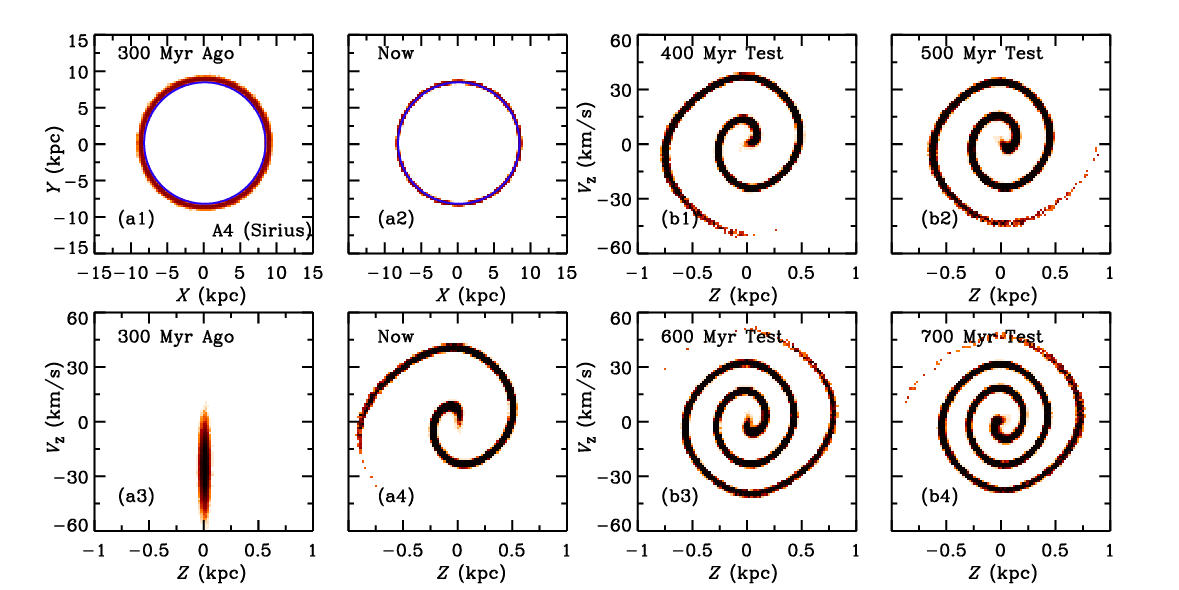}
\caption{Test particle simulation of the second approach showing the phase mixing of stars on colder orbits in arch A4 (Sirius). The layout here is the same with Fig.~\ref{fig:test1}. The snail shell is more tightly wound for the earlier stage perturbations.}
\epsscale{1.0}
\label{fig:test3}
\end{figure*}

For comparison, we perform similar test particle simulation on colder orbits in arch A4 (Sirius). The test results are shown in Fig.~\ref{fig:test3}. Clearly, the snail becomes more tightly wound with the perturbation imposed at earlier stages. Compared to the snail shell shape in Fig.~\ref{fig:snail_cle} for the colder orbits, test particle simulations with perturbations imposed $\sim 500-600$ Myr ago seem to agree well with the snail shell shape displayed by the colder orbits in Fig.~\ref{fig:snail_cle}.

The test particle simulations support our conclusions on the faster phase mixing of the hotter orbits. It seems that the vertical phase mixing should have started at least $500$ Myr ago, or there is not enough time for the snail shell pattern in hotter orbits to phase-mix away. Our result helps put tighter constraints on the vertical perturbation event of the Milky Way disk, which was suggested to occur $\sim 300 - 900$ Myr ago \citep{antoja_etal_18}. The lack of snail shell pattern can offer a new perspective, and can place important constraints on the occurrence time of the phase mixing event. 

Clearly, these test particle simulations still have limitations. For example, \citet{dar_wid_19} found that the phase space snail becomes less wound in a self-consistent simulation with self-gravity than the test particle run at longer time scales ($\sim$ 1 Gyr).  In the future more self-consistent simulations are desired to better constrain the detailed perturbation history of the Galactic disk.

We have shown that the existence or the lack of the prominent phase space snail shells is connected to the dichotomy of colder and hotter orbits, and to the different arches in the $V_{R}-V_{\phi}$ phase space. Our results and explanations strongly argue against the suggestion that the phase space snail shell is only produced by the major moving groups with no evidence of the ongoing vertical phase mixing \citep{michtc_etal_19}. There are two obvious counter arguments to their suggestion. Firstly, arches A3 and A6, which do not contain any of the major moving groups, still show phase space snail shell. Secondly, the Hercules Slow branch (A9), being a major moving group, shows no prominent snail shell feature.

\subsection{$V_{R}$ and $V_{\phi}$ Color-Coded Phase Spaces}

\citet{bin_sch_18} suggested that the vertical and radial oscillation frequencies ($\Omega_{Z}$, $\Omega_{R}$) anti-correlate with $V_{\phi}$ (hence the angular momentum or the guiding radius $R_{g}$). Stars with different $V_{\phi}$ follow a narrow and sequential trend in the $\Omega_{Z}-\sqrt{J_{Z}}$ plane, leading to the formation of the snail shell in the $V_{\phi}$ color-coded phase space. On the other hand, faster $\Omega_{R}$ leads to faster change of $V_{R}$. Therefore, the $V_{R}$ color-coded phase space shears into a spiral that differs from the $V_{\phi}$ color-coded phase space spiral in the tightness of the winding \citep{bin_sch_18}. Different simulations also indicate that the coupled motions in horizontal and vertical directions lead to clear snail shells in $V_{R}$ and $V_{\phi}$ color-coded phase spaces \citep{dar_wid_19, laport_etal_19}. In particular, the bar buckling perturbation scenario predicts the most clear snail shell in $V_{R}$ color-coded phase space \citep{khoper_etal_19}.

However, for some arches in colder orbits, $V_{R}$ or $V_{\phi}$ color-coded phase spaces show no clear snail shell (see Fig.~\ref{fig:snail_cle}). It may be understandable for arches with a narrow range of $V_{\phi}$. For example, the $V_{\phi}$ range of  arch A4 (Sirius) is only 15 km/s, which is difficult to highlight snail shell (with larger $\Delta V_{\phi}$) in the $V_{\phi}$ color-coded phase space. In Fig.~\ref{fig:cold_hot}, if we combine all the stars on colder orbits, a moderate snail shell in $V_{\phi}$ (and even weaker in $V_{R}$) color-coded phase space is present. These results suggest that the coupling between the in-plane and vertical motions may be weaker for the colder orbits than previously expected. 

\subsection{The Enhanced Snail Shell in the $V_{\phi}$ Color-Coded Phase Space}

Mathematically speaking, the $V_{\phi}$ color-coded phase space in \citet{antoja_etal_18} is just the number weighted average of the azimuthal velocity of all the arches. In an ideal case, imagine that there is only one arch at some $V_{\phi}$ and a featureless background at lower $V_{\phi}$. If all the stars in this arch are arranged into a snail shell shape in $Z-V_{Z}$ number density phase space, then the combination with the smooth background at lower $V_{\phi}$ naturally results in a pronounced snail shell color-coded in $V_{\phi}$, which, in principle, has the same shape as the snail shell in the number density distribution.

\begin{figure}
\epsscale{1.2}
\plotone{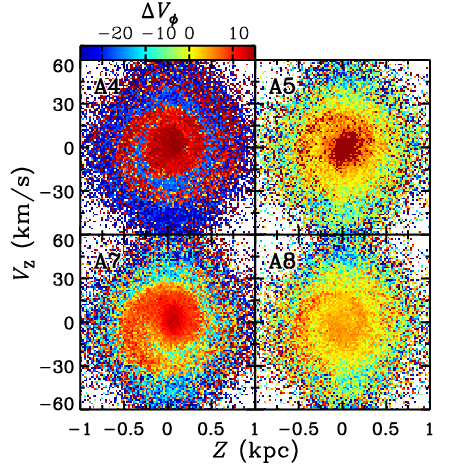}
\caption{The $V_{\phi}$ color-coded $Z-V_{Z}$ phase space of the arches in Fig.~\ref{fig:snail_cle} showing clear snail shell combined with the arches on hotter orbits in Fig.~\ref{fig:snail_no} with relatively smooth phase space distribution. The snail shell is prominent in the $V_{\phi}$ color-coded phase space after the combination, which is reminiscent to the shape in the number density phase space of each arch.}
\epsscale{1.0}
\label{fig:comb}
\end{figure}

To demonstrate this argument, we create a featureless background by combining all the stars on the hotter orbits\footnote{In order to highlight the $V_{\phi}$ color-coded snail shell in the Hercules Fast branch, the arches A1 and A2 at higher $V_{\phi}$ are excluded from the hotter orbits background construction. The combination with those two arches will result in a less prominent, but still visible snail shell in the phase space when color-coded in $V_{\phi}$.}. The background is then combined with each arch on the colder orbits, namely, A4, A5, A7, and A8. Clear snail shells can be seen in the $V_{\phi}$ color-coded phase space of each composition shown in Fig.~\ref{fig:comb}. Consistent with our expectation, for each arch in Fig.~\ref{fig:comb}, the shape of the $V_{\phi}$ color-coded snail shell is consistent with the corresponding $\Delta N$ map as shown in Fig.~\ref{fig:snail_cle}.

\begin{figure}

\center
\includegraphics[scale=1.2, viewport = 140 0 120 130]{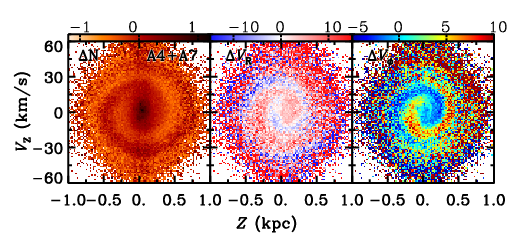}

\caption{The $Z-V_{Z}$ phase space distribution of the combination of arches A4 and A7. The left, middle and right panels show the phase space color-coded with $\Delta N$, $V_{R}$ and $V_{\phi}$, respectively. Note that the snail shell in the $\Delta N$ map is different from that in the $V_{R}$ or $V_{\phi}$ color-coded phase space.}
\label{fig:cold_comb}
\end{figure}

On the other hand, if we combine arches with different snail shell shapes at different $V_{\phi}$, then the outcome color-coded with $V_{\phi}$ (or $V_{R}$) will be a snail shell slightly different from the number density map of the combination. To test this argument, we combine arches A4 and A7 together. The result is shown in Fig.~\ref{fig:cold_comb}. Apparently, the number density contrast map and $V_{R}$ and $V_{\phi}$ color-coded phase spaces show slightly different snail shell shapes, especially in the inner region of the phase space. This is mainly due to the different snail shell shapes between the two arches. From Fig.~\ref{fig:snail_cle}, we can see that the shapes of the snail shell of the arches are not identical, especially in the central part of the phase space. Therefore, combining the arches in the colder orbits would result in different snail shell shapes between the number density map and the $V_{\phi}$ (or $V_{R}$) color-coded phase spaces, as shown in Figs.~\ref{fig:cold_hot} and~\ref{fig:cold_hot_jr}. As shown in Fig.~\ref{fig:snail_cle}, for each arch on colder orbits, the snail shell shapes are also different between the number density map and $V_{\phi}$ (or $V_{R}$) color-coded phase spaces, although the amplitude contrast of the snail shell is only $2-4$ km/s. This is probably due to the slight variation of the snail shell shape with $V_{\phi}$ (or $V_{R}$) for stars in each arch.

To summarize, it is the colder orbits that manifest the effect of ongoing vertical phase mixing, with the hotter orbits providing a featureless background to highlight the snail shell of the colder orbits in the $V_{\phi}$ color-coded phase space. Moreover, combining the arches on colder orbits could result in different snail shell shapes color-coded in $V_{R}$ or $V_{\phi}$ compared to the number density map.

\section{summary}

We provide a new perspective to understand the origin of the snail shell in the $Z-V_{Z}$ phase space and its dependence on the radial and azimuthal velocities with the Gaia DR2 data. We identify arches in the $V_{R}-V_{\phi}$ phase space, which include classical ``moving groups'' or ``kinematic streams''. Connection between the arches and the snail shell is investigated in detail. Interestingly, the snail shell only exists for stars on the colder orbits ($|V_{\phi} - V_{\rm LSR}| \lesssim 30$ km/s). Arches A3 ($\gamma$Leo), A4 (Sirius), A5 (Coma), A6, A7 (Hyades-Pleiades), and A8 (Hercules Fast) all show prominent or moderate snail shell in the phase space number density distribution, but not for the arches A1, A2, A9 (Hercules Slow), A10, A11, and A12 (Arcturus), which are mainly composed of stars on hotter orbits.

The snail shell shapes are slightly different for the arches on the colder orbits. The amplitude of the snail shell is quantified by the difference between the local maximum and minimum of the $\Delta N$ profile along a narrow $Z$ slit in the phase space. Consistent with the visual expectation, the amplitudes of the snail shell for the colder orbits ($\sim$ 0.3) are significantly higher than those of the hotter orbits ($\sim$ 0.03).

We use the radial action $J_{R}$ to quantify the extent of the radial oscillation (``hotness'') of orbits. The snail shell becomes much weaker for larger $J_{R}$ values, and essentially disappears with $J_{R} > 0.04$. Thus one should focus more on the colder orbits in the future phase mixing study, with stars on the hotter orbits removed.

We also confirm that the Hercules stream is composed of two branches with different $V_{\phi}$. Only the fast branch (A8) shows the prominent snail shell, but not the slow branch (A9). The Hercules stream may not be a homogeneous kinematic structure, which probably formed via different physical processes.

It seems that stars on hotter orbits have sufficiently phase-mixed to show significantly blurred and indistinct snail shell in the $Z-V_{Z}$ space. The hotter orbits typically have much larger dynamical range in the disk than the colder orbits. Therefore, stars on hotter orbits make blurred elliptical rotation in the $Z-V_{Z}$ phase space, which leads to faster phase mixing. These results help to put tighter constraints on the vertical perturbation history of the Milky Way disk. To explain the lack of snail shell in the hotter orbits, the Milky Way disk should be perturbed at least $500$ Myr ago.

\citet{khoper_etal_19} proposed that during the buckling process, the bar can generate bending waves in the disk to form the phase space snail shell. The bar buckling scenario predicts more pronounced snail shell in the $V_{R}$ color-coded phase space, rather than the $V_{\phi}$ or number density color-coded phase space. However, this is not seen in our results. They also suggested that the snail shell can sustain for $\sim$ 4 Gyr due to the persistence of the bending wave in the disk. It is not clear if this scenario could explain the lack of snail shells in hotter orbits. In addition, the contribution of the bar buckling mechanism on the Milky Way disk vertical perturbation may be much weaker compared to that of the Sagittarius dwarf \citep{laport_etal_19}. Also this bar buckling scenario may have some difficulties explaining recent observations on phase space distributions of different stellar populations \citep{tian_etal_18, laport_etal_19}. Apparently, more theoretical efforts are needed in the future to better constrain the time of impact and to potentially determine the mass of the perturber. The coupling between the in-plane and vertical motions for the colder orbits may be weaker than previously thought. 

The colder/hotter dichotomy in terms of the appearance of the phase space snail shell also provides a natural explanation on the significant snail shell in the $V_{\phi}$ color-coded phase space. Since only the colder orbits exhibit the effect of ongoing vertical phase mixing, the featureless phase space distribution of the hotter orbits provides a background to highlight the snail shell of the colder orbits in the $V_{\phi}$ color-coded phase space. Moreover, combining the colder orbits together could result in different snail shell shapes in the $V_{R}$ or $V_{\phi}$ color-coded phase space compared to that in the number density map.

We thank the referee for helpful comments to improve quality of the paper. We also want to thank Jerry Sellwood, Victor Debattista, Martin Smith and Chao Liu for helpful suggestions and discussions. The research presented here is partially supported by the National Key R\&D Program of China under grant No. 2018YFA0404501; by the National Natural Science Foundation of China under grant Nos. 11773052, 11761131016, 11333003; and by the ``111'' Project of the Ministry of Education. ZYL is supported by the Youth Innovation Promotion Association, and the Key Lab of Computational Astrophysics, Chinese Academy of Sciences. J.S. acknowledges support from a {\it Newton Advanced Fellowship} awarded by the Royal Society and the Newton Fund. The paper was completed at KITP, which is supported in part by NSF grant PHY-1748958. This work made use of the facilities of the Center for High Performance Computing at Shanghai Astronomical Observatory.

This work has made use of data from the European Space Agency (ESA) mission Gaia (http://www.cosmos.esa.int/gaia), processed by the Gaia Data Processing and Analysis Consortium (DPAC, http://www.cosmos.esa.int/web/gaia/dpac/consortium). Funding for the DPAC has been provided by national institutions, in particular the institutions participating in the Gaia Multilateral Agreement.


\begin{appendix}

\section{Arches Identification}

\begin{figure}
\epsscale{1.2}
\plotone{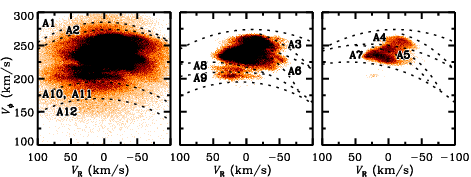}
\caption{The $V_{R}-V_{\phi}$ phase space distribution of the sample around $R = 8.34$ kpc. The three panels illustrate the identification of the gaps between the arches by changing the number density contrast levels, i.e., the minimum and maximum number densities to be displayed. Gaps at different number density levels can be much better visualized with different density contrast ranges.}
\epsscale{1.0}
\label{fig:uv_cont}
\end{figure}

The normalized empirical density distribution of the $V_{R}-V_{\phi}$ phase space is estimated with $1\times1$ (km/s)$^2$ bin size. With the wavelet transform method, \citet{ramos_etal_18} found the locations and extensions of the arches in the $V_{R}-V_{\phi}$ space. To classify stars into different arches, it is necessary to determine the gaps between these arches identified in \citet{ramos_etal_18}. Across the $V_{R}-V_{\phi}$ phase space, the stellar number density could vary up to two orders of magnitude from one arch to another, also for the gaps between the different arches. In fact, gaps between the arches with different number densities can be better visualized with different display contrast, i.e., changing the minimum and maximum number densities to be shown in the phase space. Fig.~\ref{fig:uv_cont} illustrates the $V_{R}-V_{\phi}$ phase space distribution of our sample with different number density contrast. The left panel has the lowest number density threshold (between $1.0\times10^{-6}$ and $6.2\times10^{-5}$), highlighting the arches with the lowest number density, i.e., A1, A2, A10, A11, and A12. Due to the ambient boundary between A10 and A11, they are not separated but grouped as a single arch in the analysis. In the middle panel, the lower and upper number density thresholds are increased to $4.5\times10^{-5}$ and $1.4\times10^{-4}$, respectively, leaving gaps at the higher number density visible. Arches A3, A6, A8, and A9 can be well separated. The right panel in Fig.~\ref{fig:uv_cont} highlights the three major arches in the highest density region with number density threshold between $1.0\times10^{-4}$ and $3.0\times10^{-4}$, namely, A4 (Sirius), A5 (Coma) and A7 (Hyades and Pleiades).

\end{appendix}

\end{document}